\documentclass[conference]{IEEEtran}
\IEEEoverridecommandlockouts

\usepackage{cite}
\usepackage{amsmath,amssymb,amsfonts}
\usepackage{algorithmic}
\usepackage{graphicx}
\usepackage{textcomp}
\usepackage{xcolor}
\usepackage{authblk}
\usepackage{booktabs}
\usepackage{multirow}
\usepackage{url}
\def\BibTeX{{\rm B\kern-.05em{\sc i\kern-.025em b}\kern-.08em
    T\kern-.1667em\lower.7ex\hbox{E}\kern-.125emX}}
\begin{document}

\title{Improving Perceptual Audio Aesthetic Assessment via Triplet Loss and Self-Supervised Embeddings\\
}

\author[1,2,3]{Dyah A. M. G. Wisnu}
\author[1]{Ryandhimas E. Zezario}
\author[4]{Stefano Rini}
\author[3]{Hsin-Min Wang}
\author[1]{Yu Tsao}
\affil[1]{Research Center for Information Technology Innovation, Academia Sinica}
\affil[2]{College of Informatics, National Chengchi University}
\affil[3]{Institute of Information Science, Academia Sinica}
\affil[4]{Institute of Communications Engineering, National Yang Ming Chiao Tung University}

\maketitle

\begin{abstract}
We present a system for automatic multi-axis perceptual quality prediction of generative audio, developed for Track 2 of the AudioMOS Challenge 2025. The task is to predict four Audio Aesthetic Scores—Production Quality, Production Complexity, Content Enjoyment, and Content Usefulness—for audio generated by text-to-speech (TTS), text-to-audio (TTA), and text-to-music (TTM) systems. A main challenge is the domain shift between natural training data and synthetic evaluation data. To address this, we combine BEATs, a pretrained transformer-based audio representation model, with a multi-branch long short-term memory (LSTM) predictor and use a triplet loss with buffer-based sampling to structure the embedding space by perceptual similarity. Our results show that this improves embedding discriminability and generalization, enabling domain-robust audio quality assessment without synthetic training data.
\end{abstract}


\section{Introduction}
The AudioMOS Challenge 2025 advances automatic evaluation for audio generation, with Track 2 targeting Audiobox Aesthetic Scores across four perceptual axes: Production Quality, Production Complexity, Content Enjoyment, and Content Usefulness. Participants develop models to predict these scores for TTS, TTA, and TTM systems. Building on the VoiceMOS Challenges \cite{audiomos1,audiomos2,audiomos3}, this edition extends beyond speech to broader audio content. Recent deep-learning-based non-intrusive evaluation methods\cite{yang22o_interspeech, utmos, zezario2024studyincorporatingwhisperrobust, manocha, wang2024enablingauditorylargelanguage, 10447907, mogridge2024nonintrusive, cooper2024review, fu2024selfsupervised} increasingly leverage large-scale audio pre-trained models—such as self-supervised learning (SSL) models \cite{ssl-mos,mosa-net}, Whisper \cite{zezario2024studyincorporatingwhisperrobust}, and speech language models \cite{speechlm}—to extract robust acoustic representations for improved generalization.

The challenge 2025 edition expands the scope to include music and general audio, while introducing greater domain variability. A key challenge in this track lies in the domain shift: the training and development sets consist of natural audio, while the evaluation set is composed entirely of synthetic samples. To address this mismatch, we propose a system that leverages BEATs \cite{b1}, a transformer-based audio representation model, in combination with a multi-branch architecture and a triplet loss with buffer-based sampling strategy. The triplet loss enforces a perceptually meaningful structure in the embedding space by pulling together audio samples with similar aesthetic scores and pushing apart those with dissimilar scores, improving robustness to unseen generative content.

\section{System Description}

\subsection{Dataset}\label{AA}
We used the AES-natural dataset provided by the AudioMOS Challenge 2025, which includes 2,950 natural audio samples from three domains:

\begin{itemize}
    \item Speech (950 samples from EARS, LibriTTS, and CommonVoice 13.0),
    \item Music (1,000 samples from MUSDB18-HQ and MusicCaps),
    \item General audio (1,000 samples from a subset of AudioSet).
\end{itemize}

Each audio sample is rated by 10 expert listeners across four perceptual axes: PQ, PC, CE, and CU. We use the mean of the listener scores, normalized to zero mean and unit variance, as training targets. While modeling individual listener tendencies (e.g., as in MBNet \cite{mbnet_mos} or LDNet\cite{ldnet2022}) has been shown to improve performance, we focus on the mean score for simplicity and to align with the challenge’s evaluation protocol. No listener-specific information or demographics were used. The data was split into 2,700 training and 250 validation samples. The evaluation set consists of 3,000 unseen synthetic samples from TTS, TTA, and TTM systems, which are not accessible during training.

\begin{figure*}[t]
\centering
\includegraphics[width=0.6\textwidth]{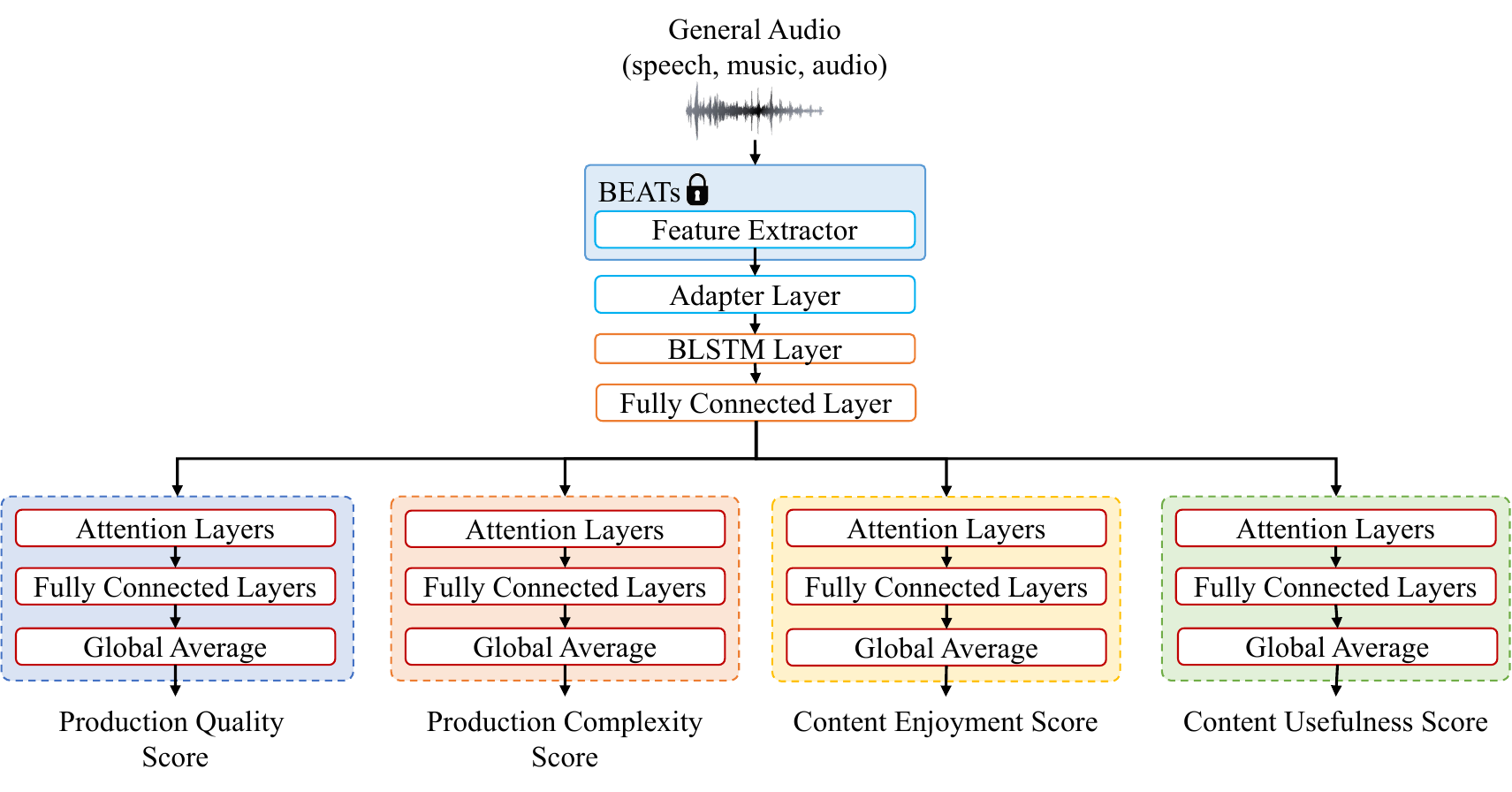}
\caption{System Architecture of AESA-Net}
\label{fig:sys_arc}
\end{figure*}

\subsection{Preprocessing and Feature Extraction}
All audio waveforms were resampled to 16 kHz. No additional preprocessing such as volume normalization or silence trimming was applied to preserve the original audio content. We use BEATs \cite{b1} (Bidirectional Encoder representation from Audio Transformers) as our self-supervised feature extractor to maintain a unified representation for both speech and non-speech audio in the evaluation data. BEATs was pretrained on AudioSet, a large-scale dataset of natural audio events, and outputs high-level representations from multiple transformer layers. Although speech-specific SSL models such as WavLM may offer more efficient representations for speech segments, integrating them with BEATs features is left as future work.



We obtain the hidden states from all BEATs layers and apply a learnable weighted sum, where each layer is scaled by a learnable scalar and the scalars are normalized via a softmax so that they sum to one. This weighted combination forms the input feature sequence for downstream modeling.

\subsection{Model Architecture}
We propose a unified multi-task architecture called \textbf{AESA-Net} (Audio Aesthetics Assessment Network), designed to jointly predict all four perceptual axes. The overall structure, inspired by HAAQI-Net \cite{b2}, is illustrated in Figure~\ref{fig:sys_arc}. The model consists of a shared backbone for feature extraction and encoding, followed by task-specific heads for each aesthetic axis.

\begin{itemize}
    \item Adapter Layer: A linear layer projects BEATs embeddings to a lower-dimensional space (linear output).
    \item BLSTM Layer: A two-layer bidirectional LSTM captures temporal dependencies from the projected features.
    \item Shared Linear Layer: Outputs from BLSTM are transformed through a shared fully connected layer followed by dropout and activation.
\end{itemize}

We then branch into four task-specific modules:

\begin{itemize}
    \item Perceptual Axes (PQ, PC, CE, CU).
    \\Each axis has its own:
    \begin{itemize}
        \item Multi-head self-attention layer (to capture content-relevant patterns), 
        \item Frame-level scoring layer (linear + sigmoid),
        \item Adaptive average pooling (to aggregate clip-level scores).
    \end{itemize}
\end{itemize}

The final outputs are four values in the [0, 1] range, representing the predicted scores for each axis. In addition, frame-level scores are available for auxiliary supervision.

\subsection{Loss Function}
We optimize a combination of mean squared error (MSE) loss and triplet loss with buffer sampling.

\section{Triplet Loss with Buffer Sampling Strategy}

To improve the generalization of the model to unseen (synthetic) audio samples, we apply \textit{triplet loss} on the intermediate embeddings using a \textit{memory buffer}. This encourages the model to learn a perceptually structured latent space, where audio samples with similar perceptual scores are close together, and dissimilar ones are pushed farther apart.

Let:
\begin{itemize}
    \item $z_a$: anchor embedding from the current sample
    \item $z_p$: positive embedding (similar perceptual score)
    \item $z_n$: negative embedding (dissimilar perceptual score)
\end{itemize}
We maintain a buffer $\mathcal{B}$ that stores recent embedding-score pairs $(z_i, y_i)$, where $y_i$ is the normalized ground-truth perceptual score (e.g., for PQ). For buffer-based sampling, we adopt the strategy proposed in \cite{b4}.

The training process proceeds as follows:
\begin{enumerate}
    \item \textbf{Buffer Update:} \\
    After processing a training sample, its embedding $z$ and corresponding target $y$ are added to the buffer $\mathcal{B}$. The buffer has a fixed capacity and behaves as a queue (FIFO), where older entries are replaced when full.

    \item \textbf{Triplet Sampling:} \\
    We follow \cite{b3}, given an anchor $(z_a, y_a)$:
    \begin{itemize}
        \item A positive example $(z_p, y_p)$ is sampled from $\mathcal{B}$ such that $|y_a - y_p| < \epsilon$
        \item A negative example $(z_n, y_n)$ is sampled such that $|y_a - y_n| > \epsilon$
    \end{itemize}
    If both are found, we compute the triplet loss:
    \begin{equation}
        \mathcal{L}_{\text{Triplet}} = \max\left( \| z_a - z_p \|_2^2 - \| z_a - z_n \|_2^2 + \text{margin},\ 0 \right)
    \end{equation}

    \item \textbf{Prediction Loss (MSE):} \\
    The model also minimizes the mean squared error (MSE) between the predicted and ground truth perceptual scores:
    \begin{equation}
        \mathcal{L}_{\text{MSE}} = \frac{1}{N} \sum_{i=1}^N \left( \hat{y}_i - y_i \right)^2
    \end{equation}
\end{enumerate}

The total loss used for backpropagation is a combination of the MSE loss and triplet loss:
\begin{equation}
    \mathcal{L}_{\text{total}} = \mathcal{L}_{\text{MSE}} + \alpha \cdot \mathcal{L}_{\text{Triplet}}
\end{equation}
where $\alpha$ is a hyperparameter (e.g., $\alpha = 0.2$) controlling the contribution of the triplet loss to the total loss.

\section{Experiments and Results}

\subsection{Training Setup}
The model was optimized using the Adam optimizer with an initial learning rate of 1e-4, and trained for up to 100 epochs with early stopping based on validation MSE. The batch size was set to 1 due to variable-length audio input and GPU memory constraints.

\subsection{Evaluation Metrics}
We evaluated the model using the official challenge metrics:
\begin{itemize}
    \item \textbf{MSE (Mean Squared Error)}: Measures the squared difference between predicted and ground truth scores for each perceptual axis.
    \item \textbf{SRCC (Spearman Rank Correlation Coefficient)}: Evaluates rank-based monotonic correlation between predicted and reference scores.
    \item \textbf{PCC (Pearson Correlation Coefficient)}: Assesses linear correlation, also referred to as LCC in some literature.
    \item \textbf{KTAU (Kendall's Tau)}: Measures ordinal association between predicted and ground truth rankings, offering a more robust alternative to SRCC in noisy scenarios.
\end{itemize}

These were computed at the \textbf{system-level} by averaging predictions and scores per sample 

\begin{table}[t]
\centering
\caption{System-level results on the official evaluation set across four perceptual axes and three domains.}
\label{tab:evaluation_results}
\begin{tabular}{llcccc}
\toprule
\textbf{Domain} & \textbf{Axis} & \textbf{MSE} & \textbf{LCC} & \textbf{SRCC} & \textbf{KTAU} \\
\midrule
\multirow{4}{*}{Speech}
& PQ & 0.5138 & 0.6950 & 0.7182 & 0.4909 \\
& PC & 0.0060 & 0.7784 & 0.5545 & 0.4182 \\
& CU & 0.9539 & 0.6520 & 0.7273 & 0.5273 \\
& CE & 1.4739 & 0.6726 & 0.7182 & 0.4909 \\
\midrule
\multirow{4}{*}{Music}
& PQ & 0.7242 & 0.8775 & 0.6703 & 0.4872 \\
& PC & 0.4843 & 0.8934 & 0.8956 & 0.7436 \\
& CU & 1.1404 & 0.8736 & 0.7198 & 0.5385 \\
& CE & 2.3145 & 0.8405 & 0.8693 & 0.6839 \\
\midrule
\multirow{4}{*}{Audio}
& PQ & 0.6514 & 0.8662 & 0.8811 & 0.7576 \\
& PC & 0.0635 & 0.8264 & 0.6154 & 0.4545 \\
& CU & 1.9407 & 0.8380 & 0.8322 & 0.6970 \\
& CE & 1.8942 & 0.8133 & 0.7483 & 0.5455 \\
\bottomrule
\end{tabular}
\end{table}

\begin{table}[t]
\centering
\caption{Comparison with the Official Baseline (System-Level)}
\label{tab:baseline_comparison}
\begin{tabular}{lcccc}
\toprule
\textbf{Axis} & \textbf{Metric} & \textbf{Baseline} & \textbf{Ours} & \textbf{Best} \\
\midrule
\multirow{4}{*}{\textbf{PQ}} 
& MSE   & 0.632 & \textbf{0.635} & $\downarrow$ \\
& LCC   & \textbf{0.903} & 0.897 &  \\
& SRCC  & 0.866 & \textbf{0.896} & $\uparrow$ \\
& KTAU  & 0.683 & \textbf{0.737} & $\uparrow$ \\
\midrule
\multirow{4}{*}{\textbf{PC}} 
& MSE   & 0.226 & \textbf{0.198} & $\downarrow$ \\
& LCC   & 0.963 & \textbf{0.984} & $\uparrow$ \\
& SRCC  & \textbf{0.934} & 0.928 &  \\
& KTAU  & \textbf{0.800} & 0.781 &  \\
\midrule
\multirow{4}{*}{\textbf{CE}} 
& MSE   & \textbf{1.142} & 3.991 &  \\
& LCC   & 0.921 & \textbf{0.950} & $\uparrow$ \\
& SRCC  & 0.841 & \textbf{0.904} & $\uparrow$ \\
& KTAU  & 0.655 & \textbf{0.741} & $\uparrow$ \\
\midrule
\multirow{4}{*}{\textbf{CU}} 
& MSE   & 1.478 & \textbf{0.533} & $\downarrow$ \\
& LCC   & 0.879 & \textbf{0.897} & $\uparrow$ \\
& SRCC  & 0.810 & \textbf{0.894} & $\uparrow$ \\
& KTAU  & 0.629 & \textbf{0.724} & $\uparrow$ \\
\bottomrule
\end{tabular}
\end{table}

\subsection{Results}

The proposed model was evaluated on the official AudioMOS Challenge 2025 evaluation set, which consists of synthetic audio samples generated from TTS, TTA, and TTM systems. Despite being trained solely on natural audio, the model was tested on three synthetic domains: speech, music, and general audio. We report the system-level performance on all four Audiobox Aesthetic Score axes—Production Quality (PQ), Production Complexity (PC), Content Usefulness (CU), and Content Enjoyment (CE)—using four standard metrics: Mean Squared Error (MSE), Linear Correlation Coefficient (LCC), Spearman Rank Correlation Coefficient (SRCC), and Kendall’s Tau (KTAU). \textbf{The complete evaluation results are summarized in Table~\ref{tab:evaluation_results}.}

The results indicate that the model generalizes well to synthetic audio, particularly in the music and general audio domains. Notably, high SRCC and KTAU values are observed for PC, suggesting strong rank-based consistency. We attribute part of this robustness to the use of triplet loss, which encourages a semantically structured embedding space and mitigates the domain gap between natural and synthetic data. Similarly, CU and CE also exhibit competitive correlation scores across domains, further demonstrating the model’s ability to maintain reliable perceptual predictions across a range of content types.

To further validate our approach, we compare our model against the official baseline system \cite{b5} provided by the challenge organizers. As shown in Table~\ref{tab:baseline_comparison}, our model achieves higher rank-based metrics (SRCC and KTAU) across all four perceptual axes, indicating improved ordinal alignment with human ratings. In particular, PQ and CU show marked improvements in SRCC and KTAU, likely benefiting from the triplet loss mechanism. Although the MSE for CE is higher than the baseline, our model yields better correlation scores (LCC, SRCC, KTAU), suggesting more accurate ranking of perceptual quality, even if the absolute values deviate. These results confirm the effectiveness of our model design and its generalization capability to unseen generative content.

\section{Conclusion}

In this work, we proposed a multi-axis perceptual quality prediction model that integrates triplet loss with a buffer-based sampling strategy to improve generalization. The model was trained solely on natural audio data but evaluated on unseen synthetic audio in the AudioMOS 2025 Challenge. Experimental results on the evaluation set show that the inclusion of triplet loss enhances both rank-based and correlation-based performance metrics. The learned embedding space exhibits improved structure and consistency across various audio domains, enabling more robust predictions even in the absence of synthetic training data.

\bibliographystyle{IEEEtran}
\bibliography{mybib}

\end{document}